\documentclass[prb,reprint]{revtex4-1} 

\usepackage{graphicx}

\begin{document}

\title{A computer-simulated Stern-Gerlach laboratory}

\author{Daniel V. Schroeder}
\altaffiliation{Present address: Department of Physics, Weber State University, Ogden, Utah 84408-2508}
\email{dschroeder@weber.edu}
\affiliation{Department of Physics, Grinnell College, Grinnell, Iowa 50112}

\author{Thomas A. Moore}
\email{tmoore@pomona.edu}
\affiliation{Department of Physics and Astronomy, Pomona College, Claremont, California 91711}

\begin{abstract}
We describe an interactive computer program that simulates Stern-Gerlach
measurements on spin-$1/2$ and spin-1 particles.  The user can design
and run experiments involving successive spin measurements, illustrating
incompatible observables, interference, and time evolution.  The program can
be used by students at a variety of levels, from non-science majors
in a general interest course to physics majors in an upper-level
quantum mechanics course.  We give suggested homework 
exercises using the program at various levels. \textsl{\copyright 1993 American Association
of Physics Teachers.}  Published in Am.\ J.\ Phys. \textbf{61} (9), 798--805 (1993), \url{<http://dx.doi.org/10.1119/1.17172>}.
\end{abstract}

\maketitle

\section{Motivation and Overview}

Quantum mechanics is central to 20th century physics, yet instructors disagree
strongly over how, and when, to teach this subject to students.
A course in quantum mechanics for beginning students faces several obstacles:  
The full theory
is horribly abstract and mathematical, while
simplified presentations tend to be vague and misleading.
The same hurdles are present, though perhaps less
severe, at the start of an upper-level quantum mechanics course for physics majors.

Much of the math can be avoided, at least for a while,
by starting with finite-dimensional spin systems.\cite{FeynmanLectures}  One
disadvantage of this approach, however, is that it makes the subject
even more abstract, as the measurable quantities are not as familiar
as position and momentum.  In principle, this problem could be
solved by assigning laboratory exercises with Stern-Gerlach devices,
but in practice such experiments are difficult and expensive to carry out.

\begin{figure}[b]
\centering
\includegraphics[width=8.5cm]{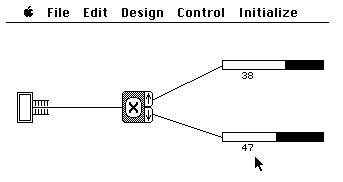}
\caption{The simplest experimental arrangement, with a single
Stern-Gerlach device and two counters.  (This experiment is ready
to run when the program starts.)}
\end{figure}

\begin{figure}[t]
\centering
\includegraphics[width=8.5cm]{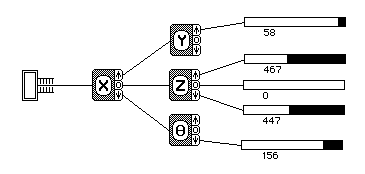}
\caption{A much more complicated experiment, involving successive
Stern-Gerlach devices and spin-1 particles.}
\end{figure}

This article describes a computer program, called {\it Spins},
that is designed to address these issues.  A Stern-Gerlach
laboratory is simulated on the computer screen (see Fig.~1), allowing the student
to quickly design and run a number of experiments involving spin systems.
As the experiment runs, simulated particles are sent through the devices one
at a time, while the student watches the numbers on the counters increase.
Multiple Stern-Gerlach devices, oriented in various directions,
can be linked together in any order, and can be used to
study spin-1 as well as spin-$1/2$ particles (Fig.~2).  Interference can
be studied by combining the beams emerging from one device and sending
them into a second (Fig.~3).  Time evolution can also be observed,
using a device that simulates the effect of a uniform magnetic field
(Fig.~4).

\begin{figure}[b]
\centering
\includegraphics[width=8.5cm]{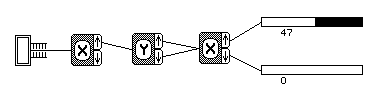}
\caption{An interference experiment with spin-$1/2$ particles.}
\end{figure}

\begin{figure}[t]
\centering
\includegraphics[width=8.5cm]{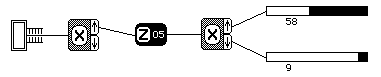}
\caption{In this experiment, spin-1/2 particles with spin up
in the $x$ direction are placed in a uniform magnetic field,
oriented in the $z$ direction, for 5 time units.  Then the
$x$ components of their spins are measured.}
\end{figure}

The pedagogical values of the program are numerous.  Most obviously, 
it gives students a concrete visual image to go with each
of the concepts just mentioned.  In addition, the program makes the
statistical nature of quantum-mechanical predictions obvious.
More generally, through designing and running their own experiments,
students see the theory at work in a somewhat realistic setting, and
learn to separate the arbitrary mathematical conventions from the unambiguous
physical predictions.

Students at almost any level can use the program.  Anyone with
sufficient curiosity can play with it and try to invent a theory to
explain the strange behavior of the particles.  Students with some knowledge
of geometry can arrive at a complete understanding
of a restricted set of experiments with
spin-$1/2$ particles.  Introductory physics
students can learn enough about complex numbers to understand
spin-$1/2$ quantum mechanics in general, while upperclass 
physics majors can explore the full complexities of a nontrivial
three-dimensional system.

Most of the rest of this article consists of guidelines and exercises written for 
users of the program.  Since the program can be used at so many different
levels, we are immediately faced with the question of who the users are.
Sections II and~III are written for students with no prior knowledge of physics 
and no mathematical background beyond plane geometry; Section~II contains
basic instructions, while Section~III contains theoretical explanations and 
suggested homework exercises at this level.  Section~IV contains a brief
discussion of uses for the program at more advanced levels, followed by
an extensive set of advanced exercises.  In Section~V we add a few 
general comments for instructors, and briefly discuss the current implementation
of {\it Spins\/} on the Macintosh.

\section{Basic Instructions}

\subsection{Getting started}

Welcome to {\it Spins}, a quantum physics laboratory at your fingertips!
When you start the program you 
see a diagram of a simple physics experiment (see Fig.~1).
At the left is a device 
(we'll call it a {\bf gun}) that emits particles (call them {\bf atoms}), 
one at a time.  The atoms follow the line to the right, then enter 
another device, an {\bf X-analyzer}, which deflects them either up or down.  
They leave the $X$-analyzer through one 
of two possible holes on its right side, then follow the paths  shown 
into two {\bf counters}.  Physicists say that this experiment ``measures''
a property of the atoms called ``spin in the $x$ direction''.  Atoms coming 
out of the upper hole of the $X$-analyzer are said to have ``spin up in the 
$x$~direction'', while those coming out of the lower hole are said to have 
``spin down in the $x$~direction''.  For this particular type of atoms, only 
these two outcomes of the measurement are possible.

To start the experiment, choose `Go' from the `Control' menu.\cite{UINote}
Let it run for a while, then choose `Stop' from the same menu.  You can start 
and stop as many times as you like.  You can reset the counters and start 
over by choosing `Reset'.  If you want to run the experiment for a long 
time but don't want to wait so long, choose `Do 1000' or `Do 10000', and 
the computer will send that many atoms through the apparatus very quickly, 
updating the counters when it is done.

You should see about half of the atoms ending up in each counter.  
Repeat the experiment several times, and convince yourself that 
although the numbers on the two counters are hardly ever exactly 
equal, the deviations from equality are in some sense ``small" and 
``random".  A physicist would say that these atoms are equally probable
to be found with their spins up or down in the $x$~direction.

This particular experiment gets tiresome before long, but you can create
different (and much more interesting) experiments in several ways.
First, put the cursor over the `X' on the $X$-analyzer and click.
The `X' will change to a `Y', and you have magically changed the 
$X$-analyzer into a new type of device, a $Y$-analyzer.  Clicking again
changes it into a $Z$-analyzer, then a $\theta$-analyzer, and finally back
into an $X$-analyzer.  Run the experiment again with the $Y$-analyzer
(and the others if you wish), and see what happens.

To make a more complicated experiment you can send the atoms
through two analyzers in succession.  
You can create more analyzers with the `New Analyzer' command
under the `Design' menu.  You can create more counters in a similar
way.  You can then move them around the screen with the mouse.
(To move an analyzer, place the cursor in the gray region surrounding its letter;
the cursor will take the form of crossed arrows.  Press the mouse button,
and drag the analyzer to its new location.)   To draw lines 
connecting the components together, first press in the output end 
of one (the cursor will take the form of an arrow pointing to the 
right), then drag to the other and release.  If you decide that you 
no longer need a component, you can select it (by clicking on it) 
and choose `Delete' from the `Design' menu.

Try designing some experiments of your own now, to get a feel for
how to manipulate the components, and to learn more about how these
atoms behave.  Use only $X$- and $Y$-analyzers for now, to keep things
simple.  What happens when you run the atoms through two successive 
analyzers of the same type?  What if the types are different?  Keep 
looking for patterns until you think you can predict the outcome
of any experiment built out of these two types of analyzers.  Try to
formulate a set of rules that would tell someone else what the outcome
of any such experiment would be.

You may be wondering what these ``analyzers'' actually are, and how they 
would work in a real experiment.  The details don't matter, but you 
should know that a $Y$-analyzer is just an $X$-analyzer turned on its side.  
The $\theta$-analyzer is also the same apparatus, but you can turn it to 
any angle with the `Change Theta' command on the `Design' menu; an 
$X$-analyzer is the same as a $\theta$-analyzer turned to zero degrees, 
while a $Y$-analyzer is the same as a $\theta$-analyzer turned to 90 degrees.  
Try running some experiments to confirm this.  Then try some other 
angles, and see if you can find a pattern.

The only other thing you need to know about the analyzers is that they
contain no moving parts (each is actually no more than a strangely shaped
magnet surrounding the path of the atoms); each analyzer always looks 
the same no matter which of its openings the atoms come out of.  Similarly,
the atoms themselves behave identically under all circumstances except 
when they are passed through one of the analyzers.  There is no way to 
tell just by ``looking" at an atom whether it will go up or down.  Of 
course this does not necessarily mean that the atoms emitted from the 
gun are all identical; that is for you to decide, on the basis of your 
experiments.

\subsection{Interference}

Here's a good experiment to try next (see Fig.~3).  Connect the gun to an 
$X$-analyzer.
Connect the up-output of this $X$-analyzer to a $Y$-analyzer.  Then connect
both outputs of this $Y$-analyzer to a second $X$-analyzer.  Connect all
remaining outputs to counters.  Run the experiment in this configuration,
then try disconnecting one of the two paths from the $Y$-analyzer to
the final $X$-analyzer.  First disconnect one, then the other, then try
it again with them both connected.  Can you explain the results?
If the rules you formulated above do not give the correct prediction
here, try to modify them to cover this experiment as well as the others.
The phenomenon exhibited here is called ``interference" by physicists,
and is analogous to the interference of light passing
through a double slit.

Since the atoms pass through the apparatus one at a time, you may wonder
if it is possible to watch each atom as it goes through, to see which
of the two paths it takes.  You can do this, but not without modifying
the analyzers.  Select `Watch' from the `Control' menu.  This attaches
a light to each output opening of each analyzer; the light bounces off
the atoms as they pass through the opening, causing a brief flash.  Now
repeat the experiment, and see what happens.

\subsection{Additional commands}

The program has several additional features, which allow you to build
still more complicated experiments.  Here is a brief summary of the
menu commands.

The `Initialize' menu determines how the gun works.  Although the atoms
coming out of it always look the same, they will behave differently
if you choose a different ``initial state".  You can choose three
different initial states, but it's up to you to determine how they
differ.  You can also choose `Random',
which randomizes the initial states.

The `Design' menu lets you create another type of experimental device:
a ``magnet" (see Figure~4).  Magnets come in four types, $X$, $Y$, $Z$, 
and~$\theta$, just 
like analyzers.  Each magnet has just one input and one output, so 
you can direct atoms through it and then use an analyzer to determine
how they have been affected.  The two-digit number on the magnet
determines how much time the atoms spend inside (in very small units
so that even 99 units of time is not long enough to noticeably slow
down the experiment).  Try setting up the experiment shown in Figure~4,
with a $Z$-magnet between two $X$-analyzers.
Increment the time (by clicking on the number) very gradually, and
systematically investigate the magnet's effect on the behavior of the
atoms as they pass through the second analyzer.  
Caution:  Experiments with magnets can be quite complicated.  To understand
the effect of magnets other than type~$Z$ requires a substantial amount
of mathematics.

You can experiment with a completely different type of atoms by choosing
`3-State Spin' from the `Design' menu.  When these atoms are sent through
an analyzer they are found to bend in one of three different directions, so the 
analyzers are now provided with three output openings instead of just 
two.  The numerical results of the experiments get much more interesting
now, but with a bit of work you should still be able to find some patterns
in the results (at least for
$X$ and $Y$, and $Z$).  Using magnets and $\theta$-analyzers in conjunction with 
the 3-state atoms is very interesting, but not recommended for beginners.

\section{Elementary Explanations and Exercises}

\subsection{The rules of quantum mechanics}

Please don't read on until you have performed several experiments
and formulated your own set of rules for predicting how the atoms
behave.  You may be able to come up with a better set of rules than
the ones given here.

Now that you've tried to understand the results of several experiments,
you may wonder how physicists understand them.  The short answer is,
we don't.  The best we've been able to come up with is a fairly concise
set of ``rules'' for calculating the (average) outcome of such an experiment.
You can decide whether these rules shed any light on what is really
going on.

The rules stated here are sufficiently general to cover experiments
involving $X$, $Y$, and $\theta$-analyzers, as well as $Z$-magnets, for the
2-state spin system.  The generalization to other experiments requires
more mathematics, but the concepts are essentially the same.\cite{ClassNote}

Rule 1:  Physicists represent the ``state" of an atom at any given time by
an arrow\cite{ArrowsVsSpinors}
drawn on a piece of paper (see Fig.~5).  All allowable arrows have 
the same length (which we'll take to be one ``unit''), but they can point
in different directions.
Arrows that point in opposite directions ($180^\circ$ apart)
represent the same physical state. 
Thus there are infinitely many possible states, 
corresponding to the infinitely many directions (from $0^\circ$ to $180^\circ$)
in which the arrow can point.

\begin{figure}[t]
\centering
\includegraphics{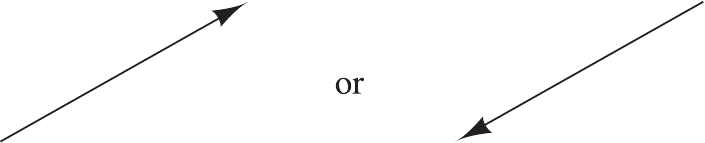}
\caption{We represent the state of an atom by an arrow drawn
on a piece of paper.  The length of the arrow is
one ``unit'', while the direction of the arrow depends on the
state in question.  Arrows pointing in opposite directions represent
the same state.}
\end{figure}

\begin{figure}[b]
\centering
\includegraphics{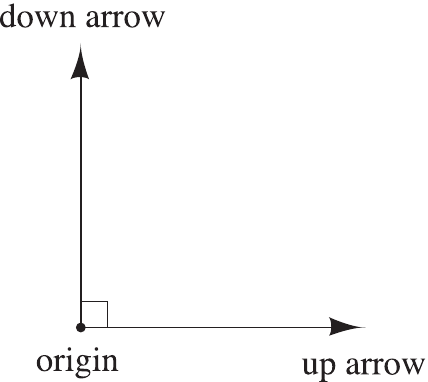}
\caption{An analyzer corresponds to a pair of unit-length arrows,
perpendicular to each other.  (The directions of the arrows depend on
which analyzer they represent.)}
\end{figure}

Rule 2:  Similarly, we represent each analyzer by a pair of 
perpendicular, unit-length arrows, whose tails coincide
(see Fig.~6).  One of these arrows corresponds to the result ``up" 
for that analyzer, while the other arrow corresponds to the result ``down".  
We will call the arrows the {\bf up arrow} and the {\bf down arrow}.  (Note
that these arrows need not, and usually do not, point up or down on 
the sheet of paper.  In general, the direction of an arrow on the paper
has no direct relation to the physical orientation of the analyzer.)
The point where the tails of the up and down arrows meet is called the {\bf origin}.

Rule 3:  When an atom passes through an analyzer, it has a certain 
probability of going up, and a certain probability of going down.
We usually can't predict which it will do;
we can only calculate the probabilities.  To compute the probability 
of going up, you first draw the arrow corresponding to the atom's current
state, and the two arrows corresponding to the analyzer, on a single
piece of paper with all their tails together at the origin (see Fig.~7).  
Suppose first that the angle between the state arrow and the up arrow is 
less than $90^\circ$.
You then draw a new line from the tip of the state
arrow, which meets the analyzer's up arrow at a right angle.  
Measure the distance from the origin to this perpendicular 
intersection of the new line and the up arrow.  This distance is called
the {\bf amplitude} for the atom to go up.
The probability for
the atom to go up is the {\it square\/} of the amplitude.  If the angle between
the state arrow and the up arrow is greater than $90^\circ$, you must first
rotate the state arrow by $180^\circ$, then follow the same procedure; in
this case the amplitude is the negative of the distance from the origin to
the perpendicular line (but the probability, being the square of the 
amplitude, is still positive).
The probability
to go down is computed in a similar way, using the analyzer's down 
arrow.  Since the state arrow has unit length, the Pythagorean theorem 
guarantees that the two probabilities add up to~1.

\begin{figure}[t]
\centering
\includegraphics{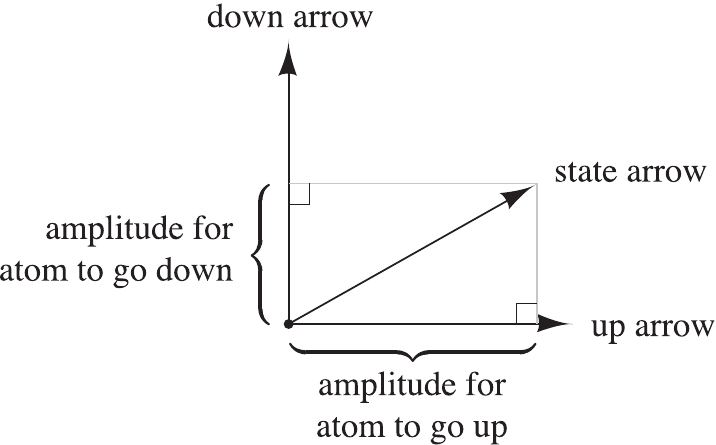}
\caption{To determine the amplitudes for the atom to go up and down, use
this geometrical construction.}
\end{figure}

Rule 4:  When the atom leaves the analyzer, its arrow changes abruptly.  If it 
went up, its new arrow is the same as the analyzer's up arrow; if it
went down, its new arrow is the same as the analyzer's down arrow.

Rule 5:  In experiments where beams are recombined (as in Fig.~3), we
must use more care in converting amplitudes to probabilities.  Consider
each path that the atom could take in order to produce a certain final
outcome, and compute the amplitude for each path by {\it multiplying\/} together
the amplitudes for each step along the path.  Then compute a total amplitude
for the final outcome by {\it adding\/} together the amplitudes for all paths.
The probability of the outcome is the square of this total amplitude.
(When there is only one possible path, this method gives the same result
as computing the probability separately for each step, as described
in Rule~3.)

Rule 6:  A $Z$-magnet causes the atom's arrow to rotate in the plane of the paper 
at a uniform speed, as long as the atom is inside the magnet.

You may have noticed that these rules are extremely abstract.  They give
only a general framework for describing quantum mechanical experiments,
without any specific prescriptions for {\it which\/} arrows we should
associate with which states.  That is because the specifics are, to a
certain degree, arbitrary.  The rules assert that we can come up with
some set of arrows to associate with the various states and analyzers,
and that once we do so, the outcomes of all experiments will be as
predicted.  The following exercises should clarify where the arbitrariness
ends and the predictions begin.

\subsection{Exercises with the 2-state spin system}

Now that you know the rules, you should be able to work the following
exercises.  Once you've done so, you will understand this system as
well as any physicist.

Figure 8 shows the up and down arrows of the $X$-analyzer.  The directions
of these arrows have been chosen arbitrarily, subject to the constraint
that they are perpendicular to each other.  Also shown in the figure
is a circle whose radius is 1~unit;\cite{FigureSize}
all other arrows, whatever they represent,
must lie with their tails at the origin and their tips somewhere on this
circle.

\begin{figure}[b]
\centering
\includegraphics{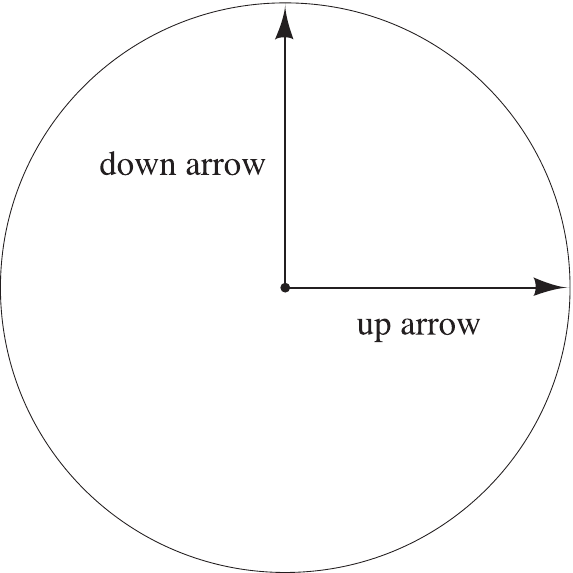}
\caption{Here we have chosen the directions of the up and down arrows of the
$X$-analyzer arbitrarily, subject to the constraint
that they be perpendicular to each other.  The circle is for reference:
all arrows must have their tails at the center and their tips
on this circle.}
\end{figure}

Set up the simple experiment shown in Fig.~1, and choose `Unknown \#3' 
from the `Initialize' menu.  By running this experiment several times,
determine the probabilities for the atoms to go up and down.  Take
the square root of each probability to determine the amplitudes for
each result, remembering that either amplitude could be positive or
negative.  Knowing these amplitudes, you can almost determine the state
arrow of the atoms:  you should be able to narrow it down to four
possibilities.  Sketch the four arrows lightly on the figure.

Add a second $X$-analyzer between the gun
and the first $X$-analyzer (see Fig.~9), and run the experiment.
Repeat the experiment with the down output (rather than the up
output) of the first analyzer connected to the input of the second. 
Also repeat the experiment with both $X$-analyzers changed to
$Y$-analyzers.
Is the behavior of this system consistent with Rule 4?  Explain
your answer carefully.

\begin{figure}[t]
\centering
\includegraphics[width=8.5cm]{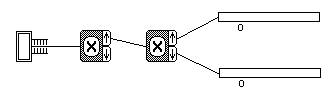}
\caption{An experiment to test Rule~4, ready to run.  (The outcome
will be that all the atoms end up in the top counter.)}
\end{figure}

Repeat the above experiment, but this time with one $X$-analyzer
and one $Y$-analyzer (in various combinations).  Using the results
of these runs,
find the directions of the up and down arrows of the $Y$-analyzer.
Again, you should find several possible directions for the arrows.
Choose one set of arrows arbitrarily, and draw them on the figure.

Now use an experiment with a single $Y$-analyzer to narrow down the choices
for the initial arrow.  
There should be two candidate arrows left, but they should point in opposite
directions.  Since Rule~1 says that arrows pointing in opposite directions
represent the same physical state, either of these arrows is correct.

Set up the interference experiment as described in Section~2.  Is this
experiment consistent with Rule~5?

Set up an experiment with a $Z$-magnet between two $X$-analyzers.
Rule~6 says that the magnet will make the atom's state
arrow rotate at a fixed rate.  By what angle does it rotate for each unit
of time spent in the magnet?   What time
setting corresponds to a full 360-degree revolution of
the arrow?  Try to design an experiment that will
tell you in which direction the arrow rotates.

\section{Guidelines for More Advanced Uses}

\subsection{Introductory physics courses}

The student instructions in Sections II and~III can easily be 
adapted for use in an introductory
physics course, where students have more mathematical knowledge and vocabulary.
The ``arrows'' can become ``vectors'', and the geometrical construction of Rule~3
can become a dot product.  This makes it possible to give purely algebraic
rules, although the geometric interpretation can still be helpful to students.
Anyone familiar with sines and cosines will have little difficulty determining the
state vectors associated with the $\theta$-analyzer.
To include rules for the $Z$-analyzer one must introduce complex numbers,\cite{BasisChoice} 
for which the geometric interpretation fails.

The time-evolution postulate (Rule 6) could be replaced by the full time-dependent
Schr\"odinger equation, although this requires some mathematical sophistication
on the part of the students.  An alternative approach in an introductory course
is to give the following algorithm for determining the time evolution of
a given initial state:  Write the initial state vector as a linear superposition
of the two orthogonal vectors associated with the direction of the magnet, then
multiply each term by a factor $\exp(-i\omega t)$, where $\omega$ equals some
constant for the ``up'' piece, and minus the same constant for the ``down'' piece.
After determining the correct constant, students can predict the outcome
of any measurement performed on atoms that have passed through a magnet.

\subsection{Upper-level physics courses}

In an upper-level quantum mechanics course for physics majors, students are
generally introduced to all the mathematical machinery of hermitian matrices,
eigenvalues and eigenvectors, and the time-dependent Schr\"odinger equation.
For the purpose of using this program, they should also be taught some version
of the postulates of quantum mechanics, analogous to those presented in the
previous section.
This allows them to deal with the 3-state system in all its complexity.
The following exercises illustrate some of the possibilities.  Many of them
are analogous to the elementary exercises of the previous section, but some
are considerably more intricate.

\subsection{Exercises for advanced students}

We will adopt the convention that all vectors are expressed in terms 
of the ``eigenbasis" of the $X$-analyzer.  This means that the eigenvectors
of the $X$-matrix are $(1,0,0)$, $(0,1,0)$, and $(0,0,1)$.  (Note that these
vectors are all normalized and mutually orthogonal.)  We will take the 
eigenvalues of the matrix to be $1$, $0$, and $-1$, corresponding to up, 
0, and down, respectively.  Associating the eigenvectors with the 
eigenvalues is also a matter of convention; we will take the
eigenvectors to correspond to the eigenvalues in the order in which
they are listed above.  Given these conventions, what is the $X$-matrix?

Choose `3-State Spin' from the `Design' menu, and `Unknown \#1' from 
the `Initialize' menu.  Use the $X$-analyzer to find the state vector
of this initial state.  You will only be able to determine the
components of the vector to within complex factors of unit modulus
(why?).  Since vectors that differ by an overall constant factor
represent the same physical state, we can choose
the convention that the first component be real and positive.  There 
are still unknown factors in the other two components, however.  Assume 
for now that the components of this vector are all real.  Then the only 
ambiguities are in the signs of the second and third components.

Hints:  All components of vectors and matrices used here can be written as
simple fractions of small integers (like $1/2$), or as square roots
of such simple fractions.  It may help you to know that if you perform
an experiment $N$ times, and the probability of a certain result is $p$,
the number of times that you actually obtain that result can differ
from the expected number $Np$ by as much as about $2\sqrt{Np(1-p)}$.  (More
precisely, the standard deviation of the distribution is $\sqrt{Np(1-p)}$;
the probability of being off by more than two standard deviations is
very small, about 5\%.)

Is the behavior of this system consistent with the ``collapse'' postulate?  
Perform successive measurements with both $X$ and $Y$ analyzers to justify your answer.

Find the eigenvalues and eigenvectors of the $Y$-matrix.  Once again you
will not be able to determine them uniquely.  Use the convention that
all components be real, and that the first component of each, as well
as the remaining components of the eigenvector corresponding to the
eigenvalue~$+1$, be positive.

These conventions for the $Y$-matrix are sufficient to determine the
unknown signs in the components of the initial state.  What are they?

Prove the identity $Y = MXM^\dagger$, where $X$ and $Y$ are the 
matrices corresponding
to the $X$ and $Y$ analyzers, $M$ is a matrix whose columns are the 
eigenvectors
of $Y$, and $M^\dagger$ is the conjugate transpose of $M$.  
This identity gives you
an easy way to find the $Y$-matrix (it can also be found by brute force,
by solving a system of linear equations).  What is the $Y$-matrix?

The $\theta$-analyzer is just an $X$-analyzer rotated by an angle $\theta$.
(A $Y$-analyzer is a $\theta$-analyzer with $\theta$ = 90 degrees.)  This 
means that the quantity measured by $\theta$ can be expressed as
$X\,\cos\theta + Y\,\sin\theta$.  The postulates say that the $\theta$-matrix
is given by this same function of the $X$ and $Y$ matrices.  What is
the $\theta$-matrix?  What are its eigenvalues and eigenvectors?
Connect one of the outputs of an $X$-analyzer to the input of a 
$\theta$-analyzer, and calculate the probabilities for obtaining the three 
possible outcomes of the $\theta$-measurement, as a function of $\theta$.
Verify your predictions by running the experiment for several different
values of~$\theta$.

A magnet causes the state vector to evolve according to the 
time-dependent Schr\"odinger equation.
The solution of this equation can be written formally as
$\psi(t) = \exp(-iHt)\psi(0)$.  Normally this expression is not very useful,
since it is usually impossible to evaluate the exponential of the matrix.  
But since we are working with very simple $3\times3$ matrices, we {\it can\/}
use this expression directly.  The matrix $\exp(-iHt)$ is called
the {\it propagator}, and denoted by $U(t)$.  We will try to determine $U(t)$
(and hence $H$) for the $Z$-magnet in the 3-state system.

Since we are using the eigenvectors of the $X$-matrix as our basis,
the matrix elements of $U(t)$ are found most easily by sending atoms
that are in $X$-eigenstates into the magnet, then measuring $X$ again
when they come out.  So connect the gun to an $X$-analyzer, one
output of this analyzer to a $Z$-magnet, the output of the magnet
to a second $X$-analyzer, and all three outputs of this analyzer
to counters.  Choose any initial state that gives you nonzero 
counts.  Increment the time on the magnet slowly, and record the
number of atoms in each counter after a reasonably long run (1000
atoms or so) for each value of the time.  Graph the relative
probability for ending up in each counter as a function of time.
Repeat the whole process using each of the three outputs of the
first $X$-analyzer.

You should now have nine graphs.  For what value of the time does 
the atom have the same state vector coming out of the magnet that 
it had going in?  (That is, for what value of $t$ does $U(t)=1$?)  Each 
graph represents the square of one element of the $3\times3$ matrix $U(t)$
(why?).  Try to guess the functional form of the curve for each of 
your graphs.  (Hints:  It is easier to guess the functional form of 
the square root of the probability.  You will get simple functions 
involving sines and cosines; for example, one of the matrix elements 
is $(1+\cos\theta)/2$.)  Remember that for any given experiment, the 
sum of the three probabilities you measured must be~1.  Do 
the functions you guessed satisfy this constraint?  Given the nine
probabilities, the elements of the matrix $U(t)$ are still unknown up
to factors of unit modulus.  To make your job easier, we have chosen
the matrix elements of~$H$ to be pure imaginary, and hence $U(t)=\exp(-iHt)$
is pure real.  Given this information, you now know the matrix elements
of $U(t)$ except for possible factors of~$-1$.

You can determine the nine unknown signs in $U(t)$ with very little
difficulty.  First note that $U(t=0)$ must equal~1; this determines
three of the nine signs.  Five of the remaining six can be found
by changing one or the other (you will have to do both, one at a time)
of the $X$-analyzers into a $Y$-analyzer.  Find a time setting on the magnet
that changes a pure $X$ eigenstate into a pure $Y$ eigenstate, or vice
versa.  The conventions used above for the $Y$ eigenstates will
determine the unknown signs.  The final unknown sign can be found
by expanding $U(t)$ as a power series in $t$ and recognizing the linear
term as $-iHt$.  The fact that $H$ is Hermitian gives one more relation
among the signs.  Write down your final expression for~$U(t)$.

The expression for $H$ you just obtained still contains an unknown 
overall constant, which you can make anything you want by re-defining
the standard unit of time.  (In other words, making the magnet twice
as strong has the same effect as leaving the atoms inside twice as
long.)  Use the following convention:  Let $t$ be the time variable 
that appears in the Schr\"odinger equation, let $n$ be the number 
displayed on the magnet, and let $N$ be the number on the magnet 
that corresponds to one full cycle (so that $U(n=N)=1$).  Then define 
$t = 2\pi(n/N)$.  Given these conventions, what is $H$?  As a check, 
you should find that the eigenvalues of $H$ are $1$, $0$, and $-1$.  
Finally, try to verify explicitly that $U(t) = \exp(-iHt)$.  (Hint:
One way to do this is first to pretend that $H$ is the diagonal
matrix diag$(1,0,-1)$, and find an expression for $\exp(-iHt)$ as a 
linear combination of 1, $H$, and $H^2$.  Then note that there exists
a matrix $M$ such that $M^\dagger M=1$ and 
$M^\dagger HM={\rm diag}(1,0,-1)$, and use this
fact to prove that your expression is correct even though $H$ is
not diagonal.)

It so happens that $H$ is identical to the matrix of the $Z$-analyzer.
Find its eigenvectors, and verify this by performing some experiments.

Referring back to your data for $X$-analyzers before and after
the magnet, for the case where the initial $X$-analyzer gave
the result `up', make a graph of the expectation value (i.e., weighted
average) of the final $X$-measurement as a function of time spent
in the magnet.  Verify that the expectation value is $\psi^\dagger
X\psi$, where
$X$ is the $X$-matrix and $\psi$ is the state vector of the atom as it
leaves the magnet.  Change the final $X$-analyzer to a $Y$-analyzer, 
and find an expression for the expectation value of $Y$ as a function
of time for the same initial state ($X$ up).  Verify Ehrenfest's
theorem, which says that for any observable $A$, $d\langle A\rangle/dt 
= -i\langle AH-HA\rangle$,
where $\langle A\rangle$ denotes the expectation value of 
the observable $A$ (and 
similarly for the observable $AH-HA$).  Note that if we consider only
expectation values, an atom behaves like a classical magnet,
initially pointing in the $X$-direction, spinning about its axis,
and precessing in the presence of a magnetic field that points in 
the $Z$-direction.  In other words, the expectation values of quantum 
observables behave as classical observables.

\section{Additional Comments}

The sets of example exercises given in Sections III.B and~IV.C are both too
short and two long.  They are too short in the sense that they could hardly
stand alone in this form---they would have to be amplified and adapted for the
specific needs of any particular course.  But they are also too long, since
working either set of exercises could take students as long as two weeks.
The minimum time investment required before the program's educational
benefits become significant is probably one week.
At least in a course for beginning students, this is a lot of time to spend
on the quantum mechanics of spin systems, a subject that has little practical
``use''.

Although this program simulates a quasi-realistic set of experiments,
it is certainly not intended as a substitute for real-life laboratory experience.
In particular, some limited experiments of this type can be performed
with polarized light and calcite crystals.  We hope that the real
experiments and the simulated ones can complement each other.

The current incarnation of {\it Spins\/} is written in Pascal and runs only 
on the Macintosh personal computer.  This program is currently available directly
from Daniel Schroeder.\cite{HowToGetSpins}

The source code for the Macintosh version of {\it Spins\/} is quite long (nearly
3000 lines), for two reasons.  First, the Pascal language is not ideally suited
to working with the complex numbers, vectors, and matrices of quantum theory,
so the implementation of the basic quantum mechanical rules is not very elegant.  
Second, a great deal of code is needed to implement the graphical
user interface.  Unfortunately, none of the user interface routines
are portable to other machines.  With sufficient time, however, one could
create an equivalent program for any machine with a graphical user interface.
The authors would be happy to assist anyone who is seriously interested
in undertaking such a project.\cite{HowToGetSpins}

On systems with a traditional ``command-line''
user interface, one can easily implement a more limited version of the program.
The experiment would proceed as a dialog such as the following:

\medskip
{\parskip=0pt\parindent=0pt\leftskip=10pt\obeylines\raggedright
User:  {\tt prepare 1000}
Computer:  {\tt 1000 atoms are being fired from the gun.}
U:  {\tt measure x}
C:  {\tt Results of X measurement: 476 up, 524 down.}
U:  {\tt select up}
C:  {\tt You have selected 476 atoms with X up.}
U:  {\tt measure y}
C:  {\tt Results of Y measurement: 261 up, 215 down.}
etc.

}

\medskip\noindent
Here the program merely keeps track of the current state vector of the system
(an array of two or three complex numbers), dots this
vector into the appropriate eigenvectors to determine the probabilities
of all possible outcomes, and then generates a random number between
0 and~1 for each atom to determine the outcome of the measurement.  When
the user selects a subset of the atoms for a further measurement, the current
state vector is set equal to the appropriate eigenvector, according to Rule~4.
An implementation of this sort obviously lacks visual images, and would make
interference experiments difficult or impossible, but most of the exercises
described in the previous sections could still be carried out, with similar
benefits to the student.  The earliest version of the {\it Spins\/} program
was of this type.  This version has been used in a sophomore-level class, and
students found that even this crude version helped make the postulates
of quantum mechanics seem more concrete.

\begin{acknowledgments}
One of us (DVS) would like to thank Prof.~Mike Casper for his lucid lecture
notes on the quantum mechanics of spin systems, which inspired the images
used in this program.  We are also grateful to
Michael Martin for putting the finishing touches on the program and
to Grinnell College for supporting this final portion of the work.
\end{acknowledgments}


\begin{thebibliography}{9}
\bibitem{UINote} These
instructions are written for the Macintosh implementation of the program, and
assume that the user is familiar with the rudiments of
the Macintosh user interface:  using the mouse to select menu items
and to ``click'', ``press'', ``drag'', and ``release''.  A brief explanation
of these terms should be included in the instructions for students.
\bibitem{ClassNote} The
rules stated here are of course much too terse to be absorbed in a single
reading.  We assume that this material would be presented and discussed
in class.
\bibitem{ArrowsVsSpinors} For students with a limited mathematical background, this 
geometrical representation using ``arrows'' is much more accessible than
the usual algebraic representation.  For the benefit of instructors, however,
we point out that a unit-length arrow is completely equivalent to a normalized
two-component spinor.  In the exercises in Section~III.B we will use the
spinors $(1,0)$ and $(0,1)$ to represent spin up and down in the $x$~direction,
and $(1,1)/\sqrt2$ and $(1,-1)/\sqrt2$ (in either order) to represent spin
up and down in the $y$~direction.  The eigenstates of the $Z$-analyzer would
then be represented by complex spinors, for which the geometrical picture
breaks down.  This is why we avoid $Z$-analyzers at this level.
\bibitem{FigureSize} To save space, the figure is
reproduced here at a very small scale.  In practice it is probably best to
give the student a worksheet on which the figure occupies most of a page;
a scale of 10~cm for the unit-length arrows works nicely.  The instructions
can then be made more concrete by replacing units with centimeters.
\bibitem{BasisChoice} In
these guidelines we have chosen to use a basis where the spin-$X$ operator
is diagonal, rather than $Z$ as is usually done.  For beginning students
this choice seems more natural.  All experimental outcomes are of course independent
of basis.
\end{thebibliography}
\end{document}